# CHARACTERIZATION OF CYCLIC HYGROTHERMAL SWELLING AND SHRINKAGE BEHAVIOR OF BALSA WOOD AND GFRP-BALSA SANDWICH STRUCTURES

Yuan Wu[1], Pascal Casari[2], Jamal Fajoui[3], Sylvain Fréour[4] and Mouna Bouziane[5]

[1]Nantes Université, GeM, UMR 6183, France. yuan.wu@univ-nantes.fr
[2]Nantes Université, GeM, UMR 6183, France. pascal.casari@univ-nantes.fr
[3]Nantes Université, GeM, UMR 6183, France. jamal.fajoui@univ-nantes.fr
[4]Nantes Université, GeM, UMR 6183, France. sylvain.freour@univ-nantes.fr
[5]Nantes Université, GeM, UMR 6183, France. bouzianemouna62@gmail.com

## 1. INTRODUCTION

The objective of developing bio-based composite [1] sandwich structures with greener core materials is to facilitate the decarbonization of industries such as aviation and maritime. This field of research has received increasing attention over the past decades. For example, balsa wood [2-5] has emerged as a highly promising alternative to foam or honeycomb cores, offering a lightweight, rapidly renewable and cost-effective solution. However, plant fibers such as balsa wood and flax often display high hydrophilic behavior [1-3], which could affect their mechanical performance and long-term durability. Additionally, their moisture diffusion behavior is highly dependent on the physical and chemical characteristics of the constituent materials, the structural dimensions, the ambient temperature, and the aging time, among other factors. It is therefore imperative to promote further research into the characterization of the cyclic hygrothermal aging behavior of balsa wood core sandwich structures, in particular the identification of moisture-induced strains and internal stresses [2, 4] in the skins and cores, as well as the investigation of moisture desorption processes and the associated thermal shrinkage phenomena [5-6] in wood fibers, resins and glass fibers.

Accordingly, this work employs two complete moisture absorption-desorption cycles to characterize the moisture diffusion behavior of specimens comprising three different materials, including the pure balsa wood, resin-infused balsa and balsa core sandwich structures with two Glass-Fiber-Reinforced-Polymer (GFRP) skins. The changes in moisture content and hygroscopic strains in the thickness, length and width directions of all specimens were investigated with a view to identifying the hygroscopic swelling and thermal shrinkage behaviors exhibited during long-term aging cycles.

## 2. MATERIALS AND EXPERIMENTAL PROTOCOL

In order to characterize the respective effects of composite skins and balsa cores on the cyclic moisture absorption-desorption aging behavior of GFRP-balsa sandwich structures, specimens consisting of three different materials were fabricated and tested, including the pure balsa wood (B1 and B2), balsa wood infused through liquid resin diffusion [3] process (R1 and R2) and balsa core sandwich specimens with two 3-layer GFRP laminate skins (S1, S2 and S3), as illustrated in Fig. 1. The dimensions of all specimens are identical, with a length of 200 mm, width of 20 mm and thickness of 10 mm. The mean thickness of the balsa wood is 9 mm, while that of each GFRP skin is 0.5 mm.

To investigate the historical effects of hygrothermal aging cycles on the moisture absorption and desorption behavior, two complete cycles were carried out, each with 35 days of absorption and 7 days of desorption. In each cycle, all specimens were first immersed in the distilled water at room temperature until the mass equilibrium was reached. They were then dried in a climatic chamber at a temperature of 35°C and a relative humidity of 30%. The variation in moisture content was identified according to Eq. (1) throughout all moisture absorption and desorption processes:

$$c(t) = \frac{m_t - m_i}{m_i} \times 100\% \tag{1}$$

Where $c(t)$ is the moisture content. $m_i$ is the initial mass, $m_t$ is the measured mass varying as a function of time t. The deformation of specimen length was measured using KEYENCE laser displacement sensors, while the changes in dimensions in width and thickness directions were measured through a ruler, as defined by Eq. (2):

$$d(t) = \frac{d_t - d_i}{d_i} \times 100\% \tag{2}$$

Where $d(t)$ is the hygroscopic strain. $d_i$ is the initial length/width/thickness, $d_t$ is the measured length/width/thickness at time t.

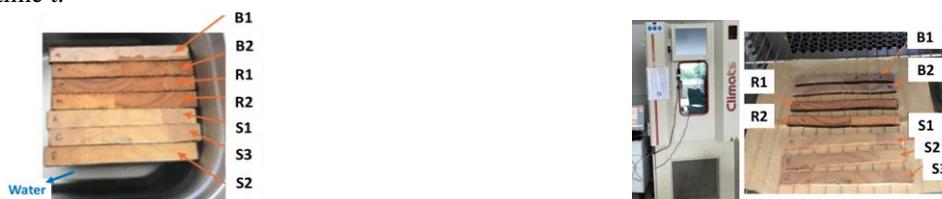

*(a) Moisture absorption test at room temperature*   *(b) Moisture desorption test in a climatic chamber*

*Fig. 1: Moisture absorption and desorption tests on balsa wood and sandwich specimens.*

## 3. EXPERIMENTAL RESULTS AND DISCUSSION

Fig. 2 presents the mean moisture content of specimens in Group B, R and S throughout the two complete absorption-desorption cycles. It is noteworthy that the repeatability of GFRP-balsa sandwich specimens is satisfactory with a maximum standard deviation of ± 6.18 %. In comparison, the pure balsa wood and resin-infused balsa specimens exhibit higher deviations of ± 49.63 % and ± 35.52 %, respectively. The final, stable moisture content capacities of the pure balsa wood, resin-infused balsa and GFRP-balsa sandwich were 499 %, 148 % and 85 %, respectively, in the initial absorption cycle, and 498 %, 138 % and 82 % in the subsequent absorption cycle. This suggests that the resin infused into the balsa core and GFRP skins has become an effective barrier preventing moisture from diffusing into the balsa core. Consequently, in the first absorption cycle, the maximum moisture content of resin-infused balsa decreased by 351 % compared to the pure balsa wood.

It is noteworthy that all specimens can be fully dried within 3 days, in both moisture desorption cycles. This observation suggests that this unique bio-based balsa wood material has a very fast moisture desorption capability and excellent associated repairability. Furthermore, the typical Fickian behavior [2] can be observed in the three groups of materials in both moisture absorption and desorption cycles.

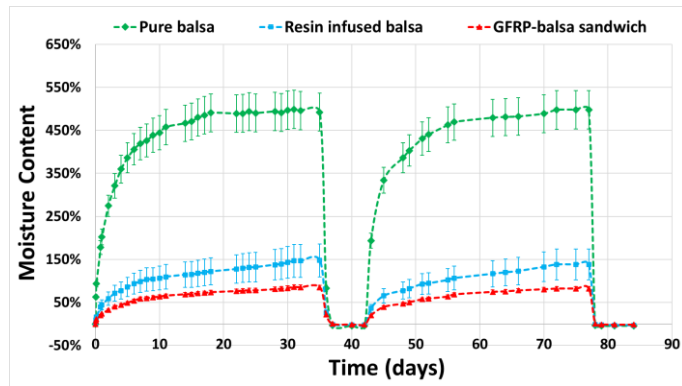

*Fig. 2: Moisture content variation in two complete moisture absorption-desorption cycles.*

The moisture-induced strains in the three directions of thickness, length and width of the specimens are presented in Fig. 3 and Fig. 4, respectively. It should be recalled that the longitudinal wood fibers of the balsa core are aligned in the thickness direction, while the length and width are considered to be the same transverse direction of anisotropic wood material [3]. Accordingly, the hygroscopic strain in the thickness direction is the predominant factor which determines the expansion effect of balsa wood structure. Fig. 3 presents that the hygroscopic strains of pure balsa wood and resin-infused balsa are almost at the same mean level, close to 4 % (as shown by solid lines), whereas it is only around 1 % in the sandwich specimens.

It can thus be concluded that the infusion of resin can result in a notable reduction of moisture content by 351 % in comparison to the pure balsa wood, while the hygroscopic strain in the thickness direction remains almost unchanged. However, the protection provided by the GFRP skins can result in a notable reduction of 3 % in moisture-induced strain in the longitudinal direction of balsa wood fiber, despite a relatively minor decrease of 63 % in moisture content compared to the resin-infused balsa. This suggests that the hygroscopic expansion in the thickness direction of GFRP-balsa sandwich can be significantly influenced by the stiffer glass fiber reinforced skins.

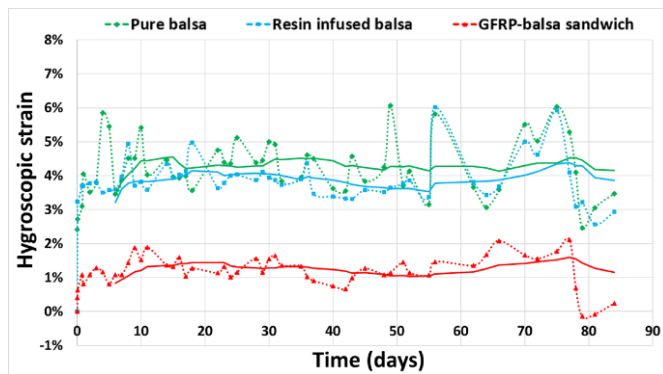

*Fig. 3: Hygroscopic strain in the thickness direction in two complete moisture absorption-desorption cycles.*

Regarding Fig. 4, it is noteworthy that the hygroscopic swelling observed during the absorption process and the thermal shrinkage evident during the desorption cycles are discernible in the length direction. However, the hygroscopic strain in the width direction appears to be relatively consistent among the three materials, with an average value of 2 %, as illustrated in Fig. 4. (b). As seen in Fig. 4. (a), it has been established that the final stable hygroscopic strains are 5.54 %, 3.30 % and 0.67 % in the pure balsa wood, resin-infused balsa and GFRP-balsa sandwich specimens, respectively, in the initial absorption cycle, and 5.16 %, 3.61 % and 0.23 % in the second resorption cycle. These results indicate a significant reduction in moisture-induced strain in the length direction of balsa wood core sandwich structure, with a decrease of 4.87 % in comparison to the pure balsa. In addition, the infusion of resin resulted in a reduction of 2.24 % in moisture-induced strain in the length direction compared to the pure balsa, suggesting that the resin and glass fibers play a significant role in controlling the hygroscopic strain in the length direction of a sandwich structure.

Another remarkable phenomenon is that negative strain values in both length and width directions can be observed in pure balsa and resin-infused balsa specimens in moisture desorption cycles, indicating that the thermal shrinkage can be triggered very quickly in the green balsa wood material under heating condition. However, this shrinkage behavior can be notably mitigated in a GFRP-balsa sandwich structure due to the protection afforded by the infused resin and outer glass fiber-reinforced skins.

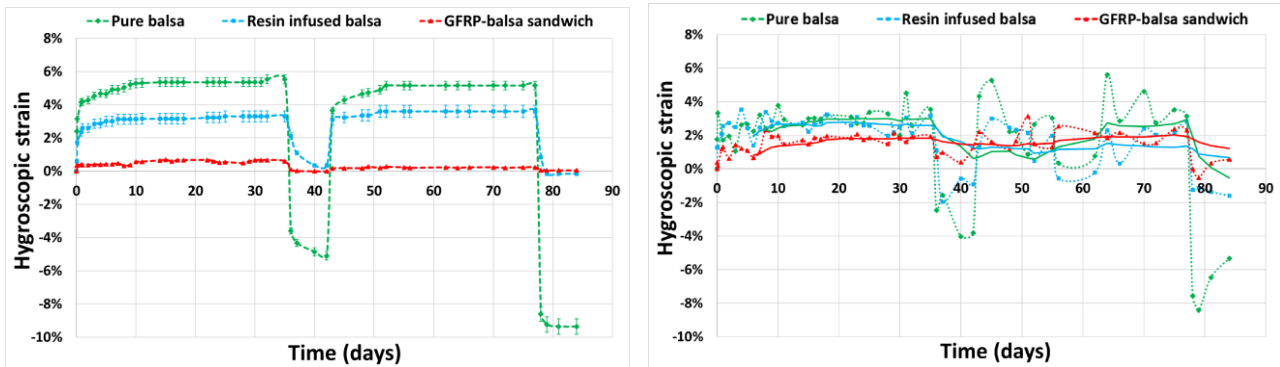

*(a) Hygroscopic strain in the length direction*  *(b) Hygroscopic strain in the width direction*

*Fig. 4: Hygroscopic strain in length and width directions in two complete moisture absorption-desorption cycles.*

## 4. CONCLUSIONS

In conclusion, this work investigates the cyclic moisture absorption-desorption behavior of pure balsa wood, resin-infused balsa and GFRP-balsa sandwich specimens over two complete hygroscopic aging cycles. The aim is to identify the different roles of balsa wood, infused resin and glass fibers in the core and skins of a bio-based sandwich structure. It has been demonstrated that the green balsa wood material possesses a high moisture absorption capacity, a fast moisture desorption capability and typical thermal shrinkage property. Nevertheless, the glass fibers in GFRP skins play a significant role in controlling the hygroscopic strain in both the thickness and length directions of sandwich structures, both during moisture absorption and desorption cycles. Moreover, the infused resin in the balsa core could significantly contribute to reducing hygroscopic swelling and thermal shrinkage along the length direction of a sandwich specimen.

Although some meaningful experimental results have been revealed in this work, further numerical analysis should be carried out on these multiple bio-based composite materials to predict the moisture diffusion kinetics in long-term cyclic hygrothermal aging cycles and to validate the contribution of different constituents to the hygrothermal aging behavior of a lightweight sandwich structure.